\documentclass[conference]{IEEEtran}
\IEEEoverridecommandlockouts
\usepackage{cite}
\usepackage{amsmath,amssymb,amsfonts}
\usepackage{algorithmic}
\usepackage{graphicx}
\usepackage{textcomp}
\usepackage{xcolor}
\def\BibTeX{{\rm B\kern-.05em{\sc i\kern-.025em b}\kern-.08em
    T\kern-.1667em\lower.7ex\hbox{E}\kern-.125emX}}

\usepackage{amsthm} 
\usepackage{bm}
\usepackage{booktabs}
\usepackage{multirow}
\usepackage{enumitem}
\usepackage{xspace}
\usepackage{url}
\usepackage{pifont}
\usepackage{flushend}

\usepackage{siunitx}
	\DeclareSIUnit\eur{\officialeuro}
	\DeclareSIUnit\M{M}
	\DeclareSIUnit\k{k}

\newcommand\ie{i.\,e.\xspace}
\newcommand\eg{e.\,g.\xspace}

\newcommand\abs[1]{| #1 |}

\newcommand{\cmark}{\ding{51}}%
\newcommand{\xmark}{\ding{55}}%

\makeatletter
\renewcommand{\fps@figure}{htb}         
\renewcommand{\fps@table}{htb}         
\makeatother 

\newcommand\pubidtext{978-1-7281-6251-5/20/\$31.00~\copyright~2020 IEEE. }
\newcommand\copyrighttext{Personal use of this material is permitted. Permission from IEEE must be obtained for all other uses, in any current or future media, including reprinting/republishing this material for advertising or promotional purposes, creating new collective works, for resale or redistribution to servers or lists, or reuse of any copyrighted component of this work in other works.}

\begin{document}
\bstctlcite{IEEEexample:BSTcontrol}

\title{
	Estimating Risk-Adjusted Hospital Performance
	\thanks{We thank the U.S. Agency of Healthcare Research and Quality (AHRQ) for providing access to the Nationwide Readmissions Database. Financial support by the Swiss National Science Foundation (SNSF) as part of the SNSF Eccellenza grant 18693 (``Data-driven health management'') is gratefully acknowledged. We further thank Oracle for computational resources.}
	\thanks{ }
	\thanks{\pubidtext \copyrighttext}
}

\author{
	\IEEEauthorblockN{Eva van Weenen}
	\IEEEauthorblockA{
		\textit{Dept. of Management, Technology and Economics}\\
		\textit{ETH Zurich, Switzerland}\\
		evanweenen@ethz.ch
	}
	\and
	\IEEEauthorblockN{Stefan Feuerriegel}
	\IEEEauthorblockA{
		\textit{Dept. of Management, Technology and Economics}\\
		\textit{ETH Zurich, Switzerland}\\
		sfeuerriegel@ethz.ch
	}
}

\maketitle

\begin{abstract}
The quality of healthcare provided by hospitals is subject to considerable variability. Consequently, accurate measurements of hospital performance are essential for various decision-makers, including patients, hospital managers and health insurers. 
Hospital performance is assessed via the health outcomes of their patients. However, as the risk profiles of patients between hospitals vary, measuring hospital performance requires adjustment for patient risk. This task is formalized in the state-of-the-art procedure through a hierarchical generalized linear model, that isolates hospital fixed-effects from the effect of patient risk on health outcomes. Due to the linear nature of this approach, any non-linear relations or interaction terms between risk variables are neglected. 

In this work, we propose a novel method for measuring hospital performance adjusted for patient risk. This method captures non-linear relationships as well as interactions among patient risk variables, specifically the effect of co-occurring health conditions on health outcomes. For this purpose, we develop a tailored neural network architecture that is partially interpretable: a non-linear part is used to encode risk factors, while a linear structure models hospital fixed-effects, such that the risk-adjusted hospital performance can be estimated. We base our evaluation on more than 13 million patient admissions across almost 1,900 US hospitals as provided by the Nationwide Readmissions Database. Our model improves the ROC-AUC over the state-of-the-art by 4.1 percent. These findings demonstrate that a large portion of the variance in health outcomes can be attributed to non-linear relationships between patient risk variables and implicate that the current approach of measuring hospital performance should be expanded. 
\end{abstract}


\section{Introduction}

Quality in the delivery of healthcare is subject to considerable heterogeneity across hospitals. To this end, some hospitals provide simply better care, confirmed repeatedly in medical research \cite{Krumholz2009b, Bernheim2010, Krumholz2017}. For instance, comparing risk-adjusted complication rates for total hip and knee replacement reveals a four-fold difference between the best and worst performing US hospitals  \cite{Bozic2014}.  
Similarly, patients at low-performing hospitals (\ie, the 25\,\% percentile) have a mortality rate after heart failure that exceeds the median mortality rate by 7.2\,\% \cite{Krumholz2009}. The underlying reasons are manifold: some hospitals have a better organizational structure, more experience in conducting certain surgeries, or offer specialized treatments. Yet, there is consistent evidence that the performance of care delivery entails considerable inter-hospital variability. 

Measuring the performance of individual hospitals is relevant for various decision-makers. Primarily, patients wish to compare hospitals based on their performance prior to choosing one. Second, managers aim at improving the quality of care, yet this is only feasible through accurate performance monitoring. Accurate performance measurements allow hospital managers to assess the effectiveness of their actions, while policy-makers can quantify the effectiveness of policy instruments and thus the overall progress towards better care.\footnote{For instance, in the US, the National Committee for Quality Assurance operates a specific program for performance improvement in healthcare called Healthcare Effectiveness Data and Information Set (HEDIS). For details, see \url{https://www.ncqa.org/hedis/.}} Third, healthcare providers such as the US Centers for Medicare \& Medicaid Services~(CMS) estimate the performance of hospitals which is then part of various downstream decision-making tasks.
For instance, measurements of hospital performance are used in the US Hospital Readmission Reduction Program to provide so-called outcomes-based reimbursement.\footnote{https://www.qualitynet.org/inpatient/hrrp} That is, financial penalties are applied to low-performing hospitals to incentivize healthcare improvement. Consequently, performance assessments are responsible for shifting financial penalties worth an estimated 563 million USD \cite{CMSFY2019}. Altogether, the previous examples highlight the need for accurate measurements of hospital performance. 

Measuring hospital performance aims at quantifying the contribution of the hospital to health outcomes of patients. Yet specifying this objective in a rigorous mathematical formalization is challenging as the base population of patients varies across hospitals \cite{Horwitz2011}. The following examples illustrate the challenge of measuring hospital performance. As specialized hospitals are visited by patients with more severe conditions, the probability of achieving a positive health outcome is lower. Similarly, hospitals located in a neighborhood with a low average income might attract patients that cannot afford specialized treatments. In the same regard, some neighborhoods have mostly elderly residents who also have overall poorer health. Given the previous examples, a simple comparison of performance indicators across hospitals is precluded. 

\textbf{Task: }
Measuring hospital performance must cater for the different risk profiles of patients and isolate their influence on health outcomes \cite{Krumholz2017}. This is formally achieved by conducting a risk adjustment \cite{Horwitz2011, YaleNewHaven2019}, where one controls for (i)~socio-demographic characteristics of patients (\eg, age, gender) and (ii)~different states of health as defined by the past medical history. Hence, the estimation task is the following: given data on patient risk profiles and their observed health outcomes $\bm{y}$, the objective is to estimate the hospital-specific contribution $\alpha_k$ to $\bm{y}$. Accordingly, the objective differs from traditional machine learning in healthcare \cite{Cheng2016, Che2016, Che2017, Ma2018}, since predicting $\bm{y}$ is not the prime interest. Instead, the estimation task requires a machine learning model that is partially \textbf{interpretable}, from which the contribution of $\alpha_k$ to the expected health outcome can be estimated.   

In current practice, the aforementioned estimation task is formalized by a hierarchical generalized linear model~(HGLM) \cite{Horwitz2011, YaleNewHaven2019}. HGLMs perform risk-adjustment by modeling the hospital-specific performance as random-effects and then isolating their values from other risk variables. Yet this approach has the following shortcomings. (1)~It is limited to linear relationships and, hence, non-linearities among risk factors are ignored. This approach is inconsistent with evidence from medical research which demonstrates that for many diseases, negative health outcomes arise both in young and old patients, thus pointing towards a non-linear relation.
(2)~Interactions between risk variables are neglected. However, medical findings stipulate that many diseases co-occur (\ie, so-called comorbidities) and that the joint effect of multiple co-occurring diseases further worsens a patient's health outcome. As illustrated in Appendix~\ref{app:data}, half of the patients visiting hospitals in the US have been diagnosed with 11 or more co-occurring diseases. Additionally, some health outcomes are the result of co-occurring patterns due to both certain socio-demographic variables \textit{and} certain prior diseases being present (\eg, the physiological responses to disease varies across genders). (3)~Current practice relies upon manual variable selection to prevent overfitting. For this reason, the HGLM is given only a small subset of available risk variables. Instead, a model would be desirable that can easily learn patterns among several thousands of different diseases.  

\textbf{Novelty: }
To address the aforementioned challenges, we propose a new non-linear model for measuring risk-adjusted hospital performance. It fulfills two requirements. On the one hand, it is partially interpretable. That is, it includes a linear component with hospital fixed-effects. This allows us to isolate inter-hospital variations in performance from risk variables, such that we can make estimates regarding the hospital-specific influence on health outcomes. On the other hand, risk variables are combined via a neural network. This flexible model allows us to learn non-linearities (including interactions) among risk variables. To the best of our knowledge, this is the first model for risk-adjustment that caters for the presence of non-linear relationships. 

Formally, our proposed model is given by a tailored neural network architecture that consists of four components: (i)~\emph{Diagnosis embeddings} provide an embedding layer that maps prior diagnoses (\ie, the disease codes from a patient's medical journey) onto lower-dimensional representations. By using pre-trained embeddings, our model can leverage the relatedness of diagnoses during learning. (ii)~In a \emph{permutation-invariant layer}, the embeddings of a variable-sized, unordered set of diagnoses are combined such that the neural network is invariant towards the ordering of diagnoses. (iii)~A \emph{fusion layer} learns a combined representation of both diagnoses and socio-demographic variables. (iv)~\emph{Hospital fixed-effects} are modeled by combining the previous output from the fusion layer with hospital dummies in a linear manner to estimate hospital performance. Note that components (i)--(iii) are non-linear and thus provide the potential for a better predictive capacity; however, component (iv)~must be linear to maintain partial interpretability and estimate the hospital performance. 

\textbf{Findings: }
Our findings are based on a US claims dataset, namely the Nationwide Readmissions Database. It is widely used by policy-makers, such as the {National Committee for Quality Assurance}, with the purpose of monitoring hospital performance and has been subject to study in prior literature \cite{Horwitz2011, YaleNewHaven2019}. The Nationwide Readmissions Database comprises claims for more than 13 million hospital admissions across 1,889 hospitals in 2016. We follow conventional guidelines in medical practice \cite{Horwitz2011, YaleNewHaven2019} where the performance of care delivery is measured by the 30-day readmission risk. Compared to the state-of-the-art HGLM, our proposed model achieves a superior model fit as demonstrated by a 4.1\% increase in ROC-AUC. 

\textbf{Implications: }
Our work has direct implications for decision-makers. To this end, our findings demonstrate that a considerable portion of the variance in health outcomes has been unexplained. Hence, the state-of-the-art approach in the form of the HGLM might erroneously attribute a superior performance to hospitals, yet only because it neglects comorbidities and other sources of non-linearities. As a result, the estimated performance in the status quo is subject to systematic errors. Here our work highlights the need for better modeling heterogeneity in multi-disease settings. This is especially relevant since, based on the estimates from HGLM, healthcare providers like Medicare assign financial penalties to US hospitals in the multi-million USD. A potential remedy is provided by the proposed model. To foster uptake in practice, the code will be made publicly available via \url{https://github.com/evavanweenen/riskadjustment}.

\section{Related Work}\label{sec:rel-work}

Machine learning in healthcare has recently received considerable traction. We provide a brief summary of prior literature according to (i)~the outcome variable, (ii)~the predictors, and (iii)~the model. For detailed reviews, we refer to, \eg, \cite{Kansagara2011}.

\textbf{Outcome variables: }
Predicted variables are usually recorded at the patient level and include various health outcomes, such as a patient's readmission risk \cite{Hosseinzadeh2013, Yu2015, Jiang2018, Maali2018, Xiao2018}, probability of mortality \cite{Makar2015}, or the onset of adverse events \cite{Farooq2011}. The former, \ie, the 30-day readmission risk, is especially common as it represents a standard indicator for monitoring quality in care (and it is thus used later analogously).  

\textbf{Predictors: }
With the ongoing digitization in healthcare, various data sources have become available for making predictions. Patient-individual data can be collected via smart sensing technology (\eg, via smartwatches or in intensive care units). 
An alternative data source is electronic health records. These collect health outcomes along the patient journey. As such, they provide a longitudinal representation for making predictions where the time component is explicitly considered \cite{Ho2014, Wang2015, Che2016, Zhang2018}. Here literature must address typical challenges that arise in practice such as sparsity, irregularity, and noise \cite{Gong2017, Che2017, Zhang2019}.
Different from that, claims data is collected at the level of individual hospitalization. When making predictions from them \cite{Hosseinzadeh2013, Maali2018}, the temporal dimension is usually replaced by a hierarchical structure, whereby diseases from the prior health trajectory are grouped into a primary diagnosis (\ie, the reason for the hospitalization) and a set of secondary diagnoses (\ie, that were present at the time of the hospitalization event, but without formalizing their temporality prior to the hospitalization). Claims data (rather than electronic health records) are collected by health providers (like Medicare) and thus present the basis for comparing performance across hospitals \cite{nrdhcup}.

\textbf{Models: }
Various machine learning models have been developed to improve the prediction performance in the context of both disease-specific characteristics and specific structures of the recorded data. For instance, if electronic health records (and not claims data) are studied, longitudinal patient trajectories are available. This type of data can be appropriately modeled via sequence learning such as long short-term memories or hidden Markov models \cite{Yoon2016, Hatt2020}. To make estimations from electronic health records, various complex neural networks have been used \cite{Cheng2016, Che2016, Che2017, Ma2018}. However, they are usually regarded to act as black-box and, owing to that, interpretability is principally precluded. Yet there is extensive effort to achieve powerful predictions with models that warrant interpretability, such as logistic regressions \cite{Caruana2015, EChoi2016, Xiao2018, Allam2019}. Our work builds upon both streams, yet it requires a tailored approach, such that the contribution of some variables is interpretable (\ie, hospital variables) as our objective is to estimate hospital performance, while others can benefit from the flexibility of a fully non-linear approach (\eg, risk factors). To achieve this, a new neural network architecture for estimating hospital performance is proposed later.  

\textbf{Risk-adjustment: }
The term ``risk-adjustment'' refers to a mathematical procedure in medical research where the contribution of some variable of interest on health outcomes is measured, while simultaneously controlling for risk factors of the base patient population \cite{Horwitz2011}. This is needed in various downstream applications, where the influence of some variables on health outcomes is attributed across varying patient cohorts \cite{Krumholz2017}. For instance, it allows comparing the so-called ``risk-adjusted'' rate of hospital admissions across different US states \cite{Finkelstein2016,Finkelstein2017} or across time \cite{Barfoot2018}. 

The default approach to risk-adjustment is simply controlling for risk factors (\ie, socio-demographics and prior diagnoses) within a linear model \cite{Feuerriegel2016, Horwitz2011, YaleNewHaven2019}. Here the industry standard that is applied by both Medicare and Medicaid is a hierarchical generalized linear model \cite{Horwitz2011, YaleNewHaven2019}. However, it cannot cater for any non-linearities and interactions among risk variables.

\section{Method}\label{sec:method}

\subsection{Problem Statement}

The objective of our work is to determine the contribution of a hospital's performance on the expected health outcome. Key to this estimation is that each hospital is visited by a different base population of patients. Hence, it is needed to control for the different risk profiles across hospitals (\ie, called risk-adjustment). 

Let us introduce the following notation for our proposed model. Let $k \in \{1, \dots, K\}$ refer to different hospitals in our setting. Each of the hospitals has a risk-adjusted performance $\alpha_k$, that is unknown and therefore needs to be determined. That is, we want to determine $\bm{\alpha} = (\alpha_1, \ldots, \alpha_{K})^\top \in \mathbb{R}^{K}$.

\textbf{Input: }
We are given data $\mathcal{H}$, which contains details on the hospital visits. That is, each visit of patient $i$ at hospital $k$ is associated with an outcome $y^i_k$. 
For patient $i$, the corresponding hospital is defined in a one-hot encoding $\bm{h}^i \in \{0,1\}^{K}$, where only $h_k^i = 1$ if patient $i$ is admitted to hospital $k$. Furthermore, each patient is linked to a vector $\bm{x}^i$ describing the patient's risk profile. The risk profile comprises three different types of covariates: 
\begin{enumerate}[leftmargin=0.5cm]
	\item \emph{Primary diagnosis:} The primary diagnosis is given by $\phi^i_0 \in \mathcal{D}$, where $\mathcal{D}$ is the set of available diagnoses. The primary diagnosis specifies the reason for hospitalization and is thus the disease that should be cured. 
	Similar to other works, diagnoses are encoded with the International Classification of Diseases~(ICD) codes \cite{WHO}. For ICD-10, this amounts to $\abs{\mathcal{D}} \approx 68,000$ different disease codes. 
	\item \emph{Comorbidities:} Each patient has potentially further health conditions that might be present. Different from the primary diagnosis, these conditions are not the prime reason for hospitalization. Instead, they are either related to the primary diagnosis (\eg, fever is often a consequence of the flu; depression often co-occurs with cancer and is thus said to be concomitant) or medically independent (\eg, somebody has a broken arm and has diabetes, which are two unrelated diagnoses). In that sense, comorbidities, or secondary diagnoses, allow us to describe a patient's past hospitalizations and thus a patient's health trajectory. Formally, the $D_i \in \mathbb{N}_0$ comorbidities for patient $i$ are given by $\left(\phi_1^i, \ldots, \phi_{D_i}^i\right) \in \mathcal{D}^{D_i}$ such that the vector $\bm{\phi}^i = \left(\phi_0^i, \phi_1^i, \ldots, \phi_{D_i}^i\right)$ describes all the patient's diagnoses. The size of the diagnoses vector $\bm{\phi}^i$ is variable and depends on the patient $i$. Within the comorbidity vector $\left(\phi_1^i, \ldots, \phi_{D_i}^i\right)$, the individual diagnoses are not further ranked, hence this vector is essentially an unordered, variable-sized set. 
	\item \emph{Socio-demographics:} Socio-demographic variables are denoted by $\bm{z}^i = (z^i_1, \ldots, z^i_M )^\top\in \mathbb{R}^{M}$. This vector contains variables such as age and gender. 
\end{enumerate}
In summary, a patient's risk profile is formalized as $\bm{x}^i = (\phi_0^i, \phi_1^i, \dots, \phi_{D_i}^i, z_1^i, \dots, z_M^i)$.
Note that the above categorization of risk variables follows current practice \cite{nrdhcup}. It is further used analogously by the HGLM approach for risk-adjustment that serves as our baseline \cite{Horwitz2011, YaleNewHaven2019}. Nevertheless, the categorization provides a generic setup: if additional covariates should be included, they are simply incorporated in $\bm{z}^i$.  

\textbf{Task: } 
In order to measure the risk-adjusted performance, the predicted health outcome $\hat{y}^i_k$ is decomposed into two parts. First, hospital fixed-effects are modeled to capture the inter-hospital variance in health outcomes. Second, an additional part adjusts for the risk profile $\bm{x}^i$ that a patient population carries. This yields  
\begin{equation}\label{eq:task}
	\hat{y}^i_k = \alpha_k + f_\theta \left( \bm{x}^i \right) 
\end{equation}
for some appropriate function $f_\theta$ parameterized by $\theta$. If $y^i_k$ is binary, then one simply models the probability of a positive outcome and applies an additional link function $\ell$ such as a logit, \ie, $\ell\left( P\left( \hat{y}^i_k = 1\right)\right)$. Later, we are interested in estimating $\alpha_k$. 

\textbf{Interpretability}: The above task demands a certain level of interpretability. To this end, an integral element in (\ref{eq:task}) is the additive structure: it allows us to isolate the fixed-effects from the risk variables, so that the risk-adjusted hospital performance can be estimated. For this reason, the conventional modeling approach where all variables are fed into some model (\ie, $\hat{y}^i_k = f_\theta \left(\bm{x}^i\right)$) is precluded. Such a model would not allow us to separate the hospital-specific contribution to $\hat{y}^i_k$, which implies that the use of many neural models from the literature \cite{Cheng2016, Che2016, Che2017, Ma2018} as baselines is prohibited.

In the above problem description, $f_\theta$ is left unspecified. If $f_\theta$ is simply set to a linear model, the overall task becomes an HGLM, analogous to current practice. However, our objective is to deliberately account for non-linearities among risk variables. Hence, we aim at a specification of $f_\theta$ that includes interaction terms and thus allows for a high degree of non-linearity. This is formally achieved by defining $f_\theta$ as a neural network in the following section. A tailored architecture is needed to accommodate a variable-sized set of comorbidities, encoded in a high-dimensional space of available disease codes, yet while allowing for partial interpretability. 

\subsection{Proposed Model} \label{sec:model}

\begin{figure}[t]
	\centering
	\includegraphics[width=\linewidth, trim={7.5mm 10mm 6mm 13mm}, clip]{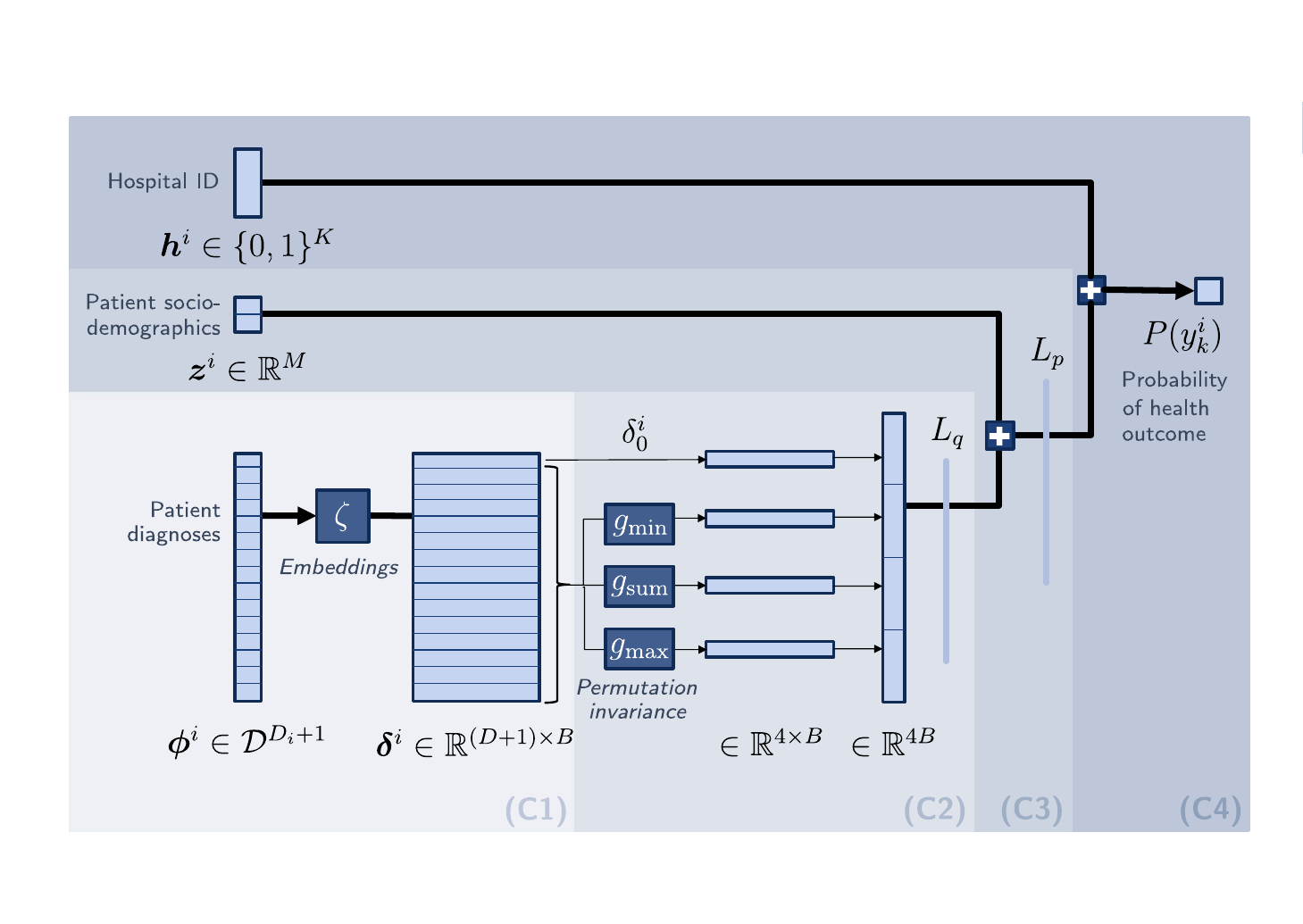}
	\caption{Proposed neural network architecture for estimating the risk-adjusted hospital performance.}
	\label{fig:nn_architecture}
\end{figure}

The proposed neural network learns the relationship between a certain health outcome $y_k^i$ of a patient $i$ admitted at hospital $k$ and patient-specific predictors $\bm{x}^i$. The latter comprises all diagnosis codes assigned to a patient and the patient's socio-demographics profile. An additional predictor is the hospital-specific contribution $\alpha_k$ towards the health outcome. This hospital-specific contribution is unknown and needs to be learned from data. Altogether, the task is addressed by the following proposed architecture, which consists of four components:
\begin{enumerate}
	\item[\textbf{(C1)}] \textbf{Diagnosis embeddings} that encode patient diagnoses into lower-dimensional representations, initialized with pre-trained diagnosis embeddings representing medical relatedness \cite{Choi2016}; 
	\item[\textbf{(C2)}] \textbf{Permutation-invariant layer} that accounts for the indifference with respect to the ordering of secondary diagnoses of a patient, implemented using the decompositions of \cite{Zaheer2017, Qi2017a, Ravanbakhsh2017}; 
	\item[\textbf{(C3)}] \textbf{Fusion layer} that combines the effects that a patient's diagnoses and socio-demographic information have on a patient's health outcome. 
	\item[\textbf{(C4)}] \textbf{Hospital fixed-effects} that combines the effects of patient diagnoses and socio-demographics with the hospital fixed-effects.
\end{enumerate}
The proposed architecture of the neural network is shown in \figurename~\ref{fig:nn_architecture}. Note that components (C1)--(C3) are non-linear and thus provide potential for a better predictive capability. In contrast, component (C4) is linear to maintain partial interpretability, so that the hospital fixed-effect can be estimated. All components are detailed in the following paragraphs.

\underline{\textbf{(C1) Diagnosis embeddings:}}
This component takes the patient diagnoses as input and maps them onto embedded vectors. Specifically, we utilize pre-trained embeddings for ICD codes that we then fine-tune. This yields a set of diagnosis embeddings $\bm{\delta}^i = \left(\delta_0^i, \delta^i_1, \ldots, \delta^i_D\right)^\top \in \mathbb{R}^{(D+1)\times B}$ for the disease codes $\bm{\phi}^i$ of patient $i$, where $B$ equals the embedding dimension, where $D = \max_i D_i $ and $\delta^i_j$ is ignored if $j > \lvert D_i \rvert$.

The rationale for embeddings in component (C1) is the following. Specifically, we opted for embeddings and discarded the use of a one-hot encoding for two reasons. First, low-dimensional, dense vectors obtained with embeddings are usually numerically beneficial over high-dimensional sparse vectors obtained with a one-hot encoding. Using embeddings thus presents a way to reduce the high dimensionality of $\mathcal{D}$. Second, diagnosis embeddings represent the medical relatedness among health conditions. A key principle of this approach is that medical concepts appearing in similar contexts should have a similar meaning. For example, pneumonia and bronchitis are more related than pneumonia and obesity, thus yielding embedding vectors in closer proximity, whereas they would receive the same distance in a binary encoding. Thereby, we can better address the (hidden) interactions between diagnoses and this approach should further facilitate learning. 

In our implementation, we use the pre-trained embeddings for the ICD, 9th Revision (ICD-9) encoding system, as obtained by \cite{Choi2016}. These embeddings were obtained using the skip-gram model implemented in the \emph{word2vec} system \cite{Mikolov2013}. The mapping of ICD codes to concept embeddings $\zeta: \mathcal{D} \to \mathbb{R}^{B}$ (here: $B=300$) obtained by \cite{Choi2016} are used as an informed initialization of our embedding layer, thus we use transfer learning to fine-tune the embeddings of the ICD codes in our algorithm. The embeddings are further processed in the subsequent components. 

\noindent
\underline{\textbf{(C2) Permutation-invariant layer:}}
This layer combines the embeddings of all diagnoses $\bm{\delta}^i$. Here the set of secondary diagnoses is processed via a permutation-invariant layer. We opted for this type of layer, rather than a recurrent neural network. The reason is that, for the secondary diagnoses, there does not exist a prior ordering by time, relevance or severity. Hence, any form of ordering is ignored with a permutation-invariant layer. In addition, this type of layer has fairly few additional parameters despite handling variable-sized input and thereby reduces the risk of overfitting. It appears furthermore suitable when considering that the number of secondary diagnoses is small (in our dataset described later, the median number of diagnoses is only 11). 

Permutation invariance is defined as follows. A function $g: \mathbb{R}^D \to \mathbb{R}$ is said to be {permutation invariant} to the ordering of its input $\{ \delta_1^i, \ldots, \delta_D^i\}$ if
\begin{equation}
	g \left(\delta_1^i, \ldots, \delta_D^i \right) = g \left( \delta^i_{\pi\left(1\right)}, \ldots, \delta^i_{\pi\left(D\right)} \right)  \quad \forall \pi \in S_D ,
\end{equation}
with $S_D$ the set of all permutations of indices $1, \ldots, D$.  
In previous literature, it was demonstrated that the permutation invariance property on a function $g$ can be achieved through decompositions of $g$ \cite{Ravanbakhsh2017,Zaheer2017,Qi2017a}. In our case, this allows us to aggregate the information from $D$ diagnoses through these decompositions.

Permutation invariance is applied to the embeddings of secondary diagnoses in our model. It is implemented through pooling functions, namely sum-decomposition, min-decomposition, and max-decomposition. The sum-decomposition is given by 
\begin{equation}\label{eq:sum-decomposition}
	g_{\text{sum}}(\delta_1^i, \ldots, \delta_D^i) = \sum_{d=1}^D {\delta_d^i}  .
\end{equation}
Min- and max-decomposition are defined as
\begin{align}
	g_{\min}( \delta_1^i, \ldots, \delta_D^i) &= \min_{1\leq d \leq D} \delta_d^i  \\
	g_{\max}( \delta_1^i, \ldots, \delta_D^i) &= \max_{1\leq d \leq D} \delta_d^i \,.
\end{align}
The embedding of the primary diagnosis $\delta_0$ and the decompositions of all secondary diagnoses are then concatenated and, consecutively, given to several dense, deep neural layers denoted by $L_q$. 
The output of $L_q$ is thus a combined representation of both primary and all secondary diagnoses. We allow $L_q$ to consist of several dense layers, such that it can learn non-linearities among patient diagnoses.

\noindent
\underline{\textbf{(C3) Fusion layer:}}
The fusion layer incorporates patient socio-demographics $\bm{z}^i$. Formally, both the output from the permutation-invariant layer (and thus the patient diagnoses) and patient socio-demographics $\bm{z}^i$ are concatenated and then jointly inserted into $L_p$. Here we allow $L_p$ to be non-linear as well, and it thus comprises several dense layers. In this way, $L_p$ learns non-linearities among both patient diagnoses and other patient covariates. 

\noindent
\underline{\textbf{(C4) Hospital fixed-effects:}} The final layer combines the output of the previous fused layer with hospital dummies. For reasons of interpretability, this is done in an additive manner, such that we can later quantify the risk-adjusted contribution of $\alpha_k$ onto the health outcomes. This yields 
\begin{equation}
	\hat{y}^i_k = \alpha_k + f_{\theta}\left(\bm{x^i}\right) = \bm{\alpha}^\top \, \bm{h}^i + f_{\theta}\left(\bm{x^i}\right),
\end{equation}
where $f_{\theta}$ denotes the neural network consisting of components (C1) to (C3).
As our outcome variable defined later is binary, an additional link function in the form of a sigmoid is used.

In our case, we aim at modeling hospital fixed-effects that follow a normal distribution, \ie,
\begin{equation}
	\alpha_k = \mu + \omega_k, \quad \omega_k \sim \mathcal{N}\left(0, \tau^2 \right) \,, 
\end{equation}
where $\mu$ is the mean of all hospital-specific effects (\ie, the so-called bias) and $\tau^2$ the variance among all hospitals. Here the objective is to extract the values of $\alpha_k$ (and, likewise, $\mu$ and $\omega_k$). This is achieved by placing an $L_2$-regularization onto the loss as follows.  

Formally, the loss is implemented as a binary-cross entropy loss with $L_2$-regularization on the weights of the hospital-specific effects $\bm{\alpha}$. This functions as a deterministic imitation of a normally distributed prior \cite{Murphy2012}, that one would have otherwise placed on the hospital-specific effect \cite{Horwitz2011}. As a result, the normal prior placed on hospital fixed-effects in the current procedure is analogously followed. Based on it, the loss function can then be written as
\begin{multline}\label{eq:loss}
	\mathcal{L} = - \sum_{i=1}^n \left[ y^i_k \log(\hat{y}^i_k) + (1-y^i_k) \, \log(1-\hat{y}^i_k) \right] \\
	+ \lambda \sum_{k=1}^K \alpha_k^2 ,
\end{multline}
where $y^i_k \in \{0,1\}$ is the health outcome of patient $i$, $\hat{y}^i_k$  the predicted health outcome of patient $i$, $n$ the total number of patients, and $\lambda$ the shrinkage factor of $L_2$-regularization.

Altogether, the above yields estimates that are partially interpretable, so that we can extract the risk-adjusted hospital performance $\alpha_k$. 

\subsection{Estimation Details}\label{sec:estimation-details}

The model is trained with the optimization algorithm Adam \cite{Kingma2015}, using cyclical learning rates \cite{Smith2017}. Further details on the training process of the neural network, as well as the hyperparameter tuning, can be found in Appendix~\ref{app:model}.

\section{Experiments}\label{sec:experiments}

\subsection{Data}

We base our numerical evaluation on the 2016 US Nationwide Readmissions Database \cite{nrdhcup}. It contains data on more than 13 million patient admissions across 1,889 US hospitals. The dataset is collected regularly by the US Agency for Healthcare Research and Quality and, as such, it is regarded as being representative of the US population. In fact, the dataset contains all patient discharge records for 27 US states, thereby accounting for around 56.6\,\% of all hospitalizations in the US \cite{nrdhcup}. The same data is also used in other studies \cite{Horwitz2011,YaleNewHaven2019}, thus ensuring comparability of our findings. Following common conventions, quality of care is measured based on the 30-day readmission rate.

The dataset provides variables on patient and hospitalization level. On patient level, this includes socio-demographics, namely gender and age (\ie, $\bm{z}^i$), as well as the patients' diagnoses (\ie, ICD codes, $\bm{\phi}^i$). The latter consist of a single primary diagnosis and a variable-sized set of secondary diagnoses. On a hospitalization level, the dataset contains the (anonymous) hospital ID, as well as patient linkage IDs through which a binary outcome variable consisting of whether a 30-day readmission occurred ($y^i$) can be created. All of the previous variables are given to our model. To this end, the choice of variables is analogous to \cite{Horwitz2011,YaleNewHaven2019}. Notably, it is governed by the current practice of healthcare policy in the US (and which reporting / privacy rules are enforced). 

Details on the pre-processing and additional summary statistics are provided in Appendix~\ref{app:data}.

\subsection{Baselines}

Our objective is to estimate the risk-adjusted hospital performance. This has important implications for the choice of our baselines. That is, any model for comparison must include an interpretable part, here a linear term modeling hospital fixed-effects, which is the reason why most approaches from the literature must be discarded. This is especially true for fully-connected neural networks \cite{Cheng2016, Che2016, Che2017, Ma2018}. The remaining approaches fulfilling the prerequisite of partial interpretability are listed as follows: 

\textbf{Hospital-mean}: This baseline represents a naive comparison in which only hospital fixed-effects are included as features, thereby omitting any form of risk-adjustment. This baseline thus allows us to quantify the marginal contribution of risk variables on the prediction performance. Formally, we simply predict health outcomes based on the hospital that a patient is admitted to, \ie, $\hat{y}^i_k = \alpha_k = \bm{\alpha}^\top \, \bm{h}^i$.

\textbf{HGLM} \cite{Horwitz2011, YaleNewHaven2019} (current practice): This baseline is given by hierarchical generalized linear models. It models hospital fixed-effects and thus fulfills the demand of being interpretable. All other risk variables are added in a linear manner. Formally, it is given by 
\begin{align}\label{eq:GLIMMIX1}
	& \hat{y}^i_k = \alpha_k + \bm{\beta}^\top \bm{v}^i = \alpha_k + \sum_{m=1}^M {\beta_m \, v^i_m} \\
	\label{eq:GLIMMIX2}
	\text{with } & \alpha_k = \mu + \omega_k , \quad \omega_k \sim \mathcal{N}\left(0, \tau^2 \right) ,
\end{align}
where $\mu$ represents the adjusted average performance over all hospitals and $\tau^2$ the inter-hospital variance (\ie, resulting in a so-called GLIMMIX model). Similar to our proposed model, the features $\bm{v}^i$ consist of patient diagnoses ($\bm{\gamma}^i$), as well as all patient-specific socio-demographics ($\bm{z}^i)$. However, rather than using high-dimensional ICD codes as patient diagnoses features, ICD diagnoses $\bm{\phi}^i$ are mapped onto the lower-dimensional CCS disease categories $\bm{\gamma}^i$, cf. \cite{YaleNewHaven2019}.

\textbf{Elastic net}: Diagnoses are likely subject to multi-collinearity and, as a remedy, shrinkage methods for estimating (\ref{eq:GLIMMIX1}) have been suggested as a robust alternative \cite{Feuerriegel2016}. Here we use $L_1$-regularization, besides the previously mentioned $L_2$-regularization, such that the estimator has the same objective as the above procedures for reasons of comparability. 

As a comparison, we also introduce a fully non-linear neural network. This allows us to quantify how much prediction performance is lost due to interpretability, \ie estimates of hospital performance. 

\textbf{Fully non-linear}: This model is an adaption of our proposed neural network architecture. The model uses the same input variables, however, it is allowed to combine both hospital dummies and risk variables in a non-linear manner. This is achieved by discarding the linear layer (C4), whereby the hospital dummies $\bm{h}^i$ are concatenated to the socio-demographic variables $\bm{z}^i$. Consecutively, the combined vector is given to the fused layer (C3). Reassuringly, we emphasize that this model does not fulfill our objective of estimating the latent hospital performance. 

\subsection{Performance Metrics}

The prediction performance and thus the model fit of the different approaches are compared across the following metrics: 
precision; recall; F1 score; ROC-AUC as the area under the receiver operating characteristic curve; and PR-AUC as the area under the precision-recall curve. The latter is common for datasets in which a large class imbalance is present.

\subsection{Numerical results}

Table~\ref{tab:results-model} compares the model fit of the different methods based on the prediction performance on out-of-sample data. Evidently, our proposed neural network demonstrates the best performance. As such, the HGLM from current practice is consistently outperformed. For instance, we advance over the ROC-AUC of the HGLM by 4.1\,\%. This performance gain can be attributed to modeling non-linearities among the risk variables. 

For reasons of completeness, Table~\ref{tab:results-model} also lists a fully non-linear baseline. This model is given identical data, yet has more degrees of freedom (\ie, it is not constrained to the interpretability of hospital fixed-effects). As such, this baseline demonstrates how much prediction performance of our proposed model architecture is lost due to interpretability constraints. We find that the performance of the fully non-linear baseline is similar to the performance of our proposed model. From this result, we conclude that no performance is lost due to interpretability constraints for our specific model architecture. Elevating interpretability constraints, however, gives rise to far more complex model architectures, that would likely obtain a higher prediction performance \cite{Cheng2016,Xiao2018,Min2019,Feng2019}.   

Additional results are relegated to Appendix~\ref{app:model}.

\begin{table}[t]
	\centering
	\scriptsize
	\caption{Out-of-Sample Prediction Performance}
	\label{tab:results-model}
	\setlength{\tabcolsep}{3pt}
	\begin{tabular}{l c c c c c c }
		\toprule
		& \textbf{PI}$^\dagger$ & \textbf{Precision} & \textbf{Recall} & \textbf{F1} & \textbf{ROC-AUC} & \textbf{PR-AUC} \\
		\midrule
		{Hospital-mean} & \cmark & 0.149 & 0.538 & 0.234 & 0.557 & 0.155\\
		{HGLM} \cite{Horwitz2011,YaleNewHaven2019} & \cmark & 0.215 & 0.647 & 0.323 & 0.701 & 0.241\\
		{Elastic net} \cite{Feuerriegel2016} & \cmark & 0.215 & 0.647 & 0.323 & 0.701 & 0.241\\
		\textbf{Proposed model} & \cmark & \bfseries 0.230 & \bfseries 0.667 & \bfseries 0.343 & \bfseries 0.730 & \bfseries 0.278\\
		\midrule
		{Fully non-linear} & \xmark & 0.229 & 0.664 & 0.340 & 0.727 & 0.275\\
		\bottomrule
		\multicolumn{7}{l}{\scriptsize $^\dagger$ \textbf{PI}: Partial interpretability to estimate hospital performance $\alpha_k$.}\\
		\multicolumn{7}{l}{
			\begin{minipage}{.4\textwidth}
				\begin{itemize}[leftmargin=*, align=parleft, labelsep=1cm]
					\item[(top)] Interpretable models that can estimate hospital performance
					\item[(bottom)] Black-box model for comparison
					\item[\textbf{bold}] Interpretable model with best performance
				\end{itemize}	
			\end{minipage}
		}
	\end{tabular}
\end{table}

\subsection{Estimated Risk-Adjusted Hospital Performance}

\begin{figure}[t]
	\centering
	\includegraphics[scale=.5, trim={0.2cm 0.2cm 0.2cm 0.2cm}, clip]{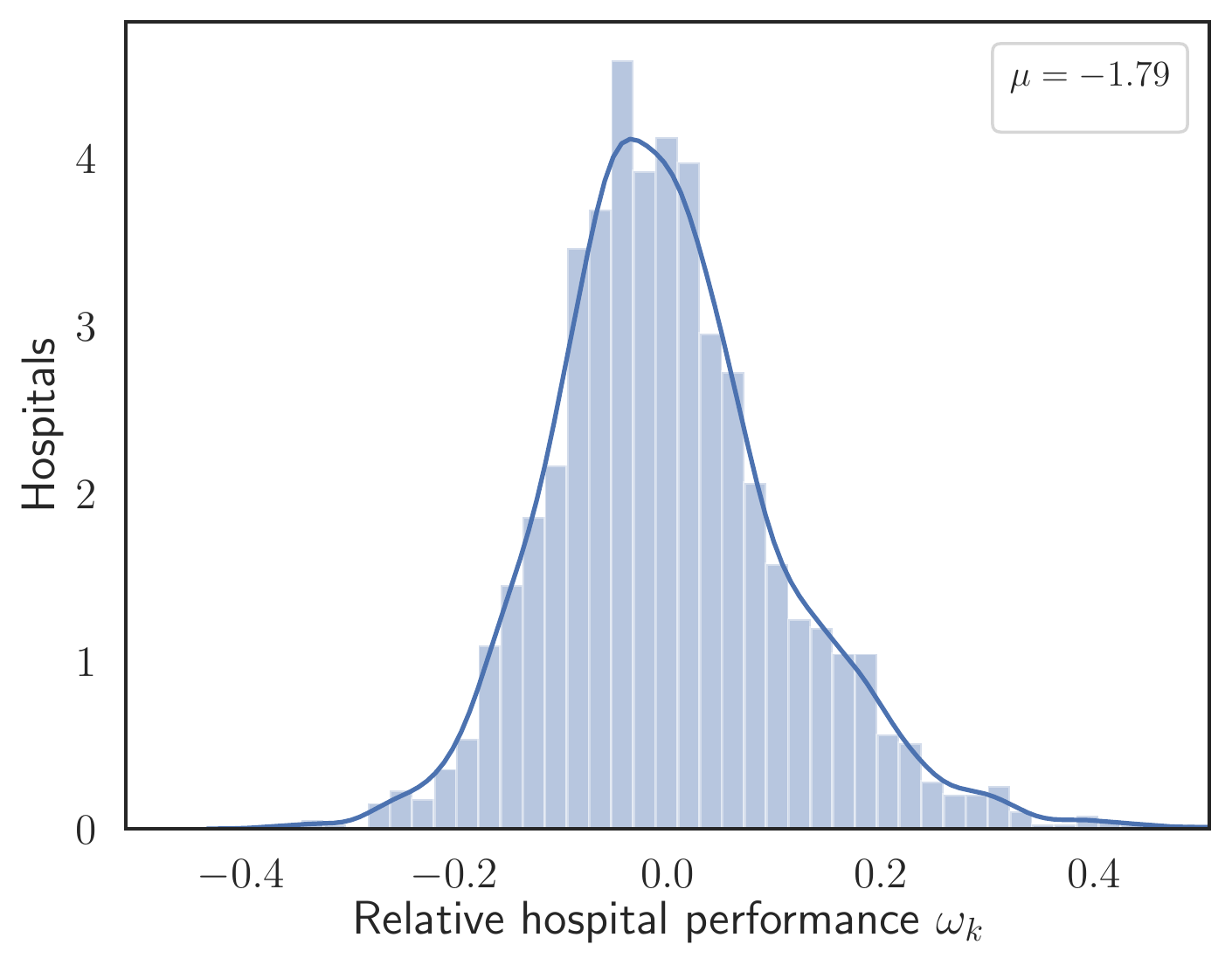}
	\caption{Estimated relative hospital-specific performance ($\omega_k$).}
	\label{fig:omegas}
\end{figure}

\begin{figure}[t]
	\centering
	\includegraphics[scale=.5, trim={0.2cm 0.2cm 0.2cm 0.2cm}, clip]{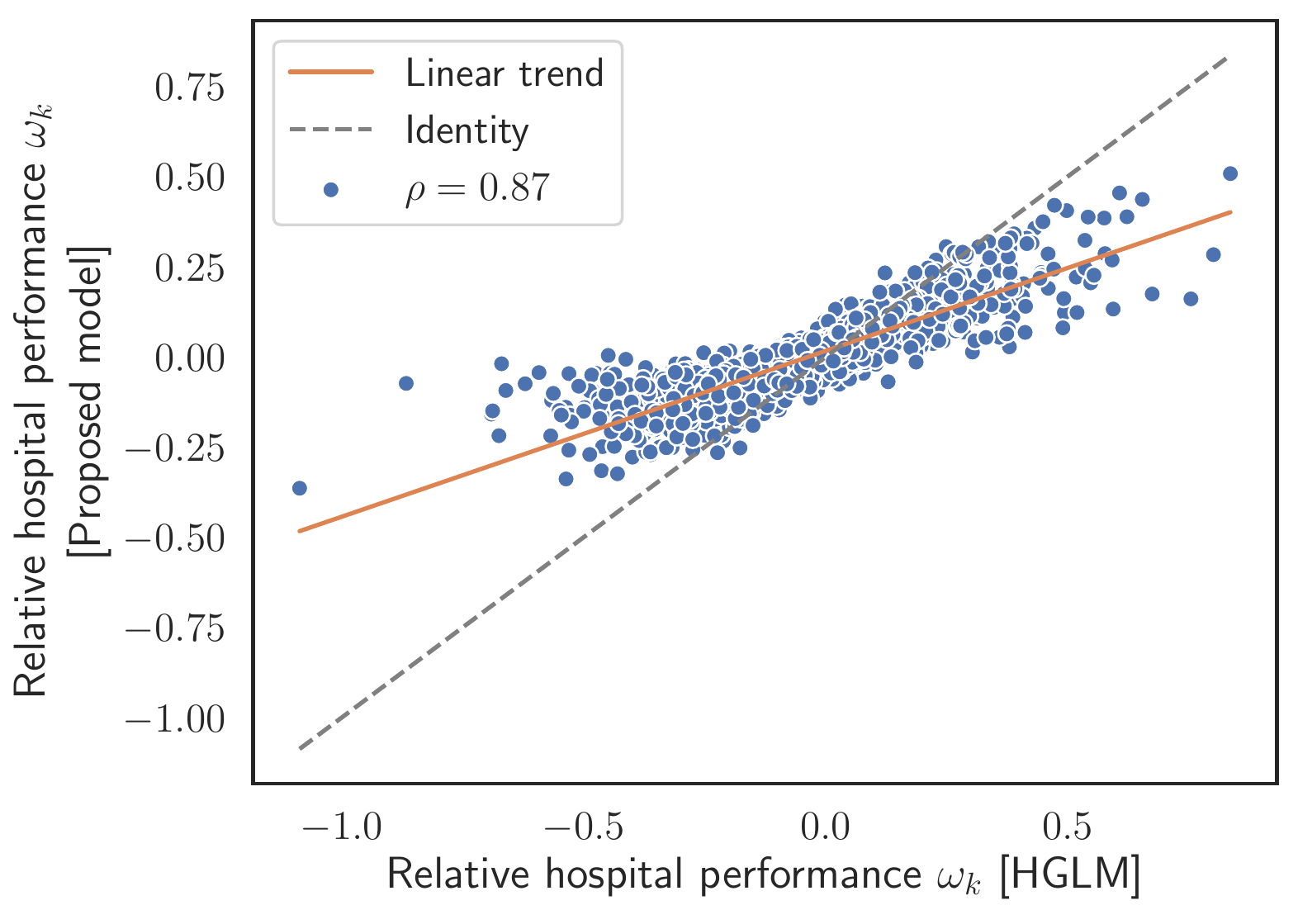}
	\caption{Comparison of hospital performance (\ie, $\omega_k$) estimated with current practice (HGLM) and with our proposed model. Shown is an identity line and a linear trend through hospital specific performances with correlation coefficient $\rho$.}
	\label{fig:omegas-comparison}
\end{figure}

With our proposed model, we can estimate risk-adjusted hospital-specific performance (\ie, $\omega_k$). \figurename~\ref{fig:omegas} shows the distribution of estimated hospital-specific performance $\omega_k$. By being fairly dispersed, the distribution points towards considerable inter-hospital heterogeneity in performance. Evidently, some hospitals have an up to 28\,\% lower or higher hospital performance $\alpha_k$ than the average hospital, where the standard deviation from the average hospital performance is 6 \%. In addition, \figurename~\ref{fig:omegas-comparison} compares the hospital performance that is estimated via (i)~risk-adjustment according to HGLM and (ii)~risk-adjustment from our proposed neural network. Overall, both appear highly correlated with few outliers. In simple words, no hospital is completely misjudged with the risk-adjustment scheme from current practice, yet there are consistent deviations for almost all hospitals. As such, the current procedure appears to make systematic errors.

\subsection{Performance across Patient Cohorts}

Table~\ref{tab:results-cohort} performs a comparison across different data splits. These are defined by \cite{YaleNewHaven2019} as patients from different cohorts. The results confirm that our model outperforms all baselines, even among different patient cohorts. Additionally, the results appear fairly stable across different cohorts (also in absolute figures). 

\begin{table}[t]
	\centering
	\scriptsize
	\caption{Prediction Performance (in ROC-AUC) Across Patient Cohorts}
	\label{tab:results-cohort}
	\setlength{\tabcolsep}{3pt}
	\begin{tabular}{l r ccccc}
		\toprule
		\textbf{Cohort}$^\ddagger$ & \multicolumn{1}{c}{\textbf{Obs.}} & \textbf{Hospital-} & \textbf{HGLM} & \textbf{Elastic} & \textbf{Proposed} & \textbf{Fully}\\
		 & & \textbf{mean} & \cite{Horwitz2011,YaleNewHaven2019} & \textbf{net} \cite{Feuerriegel2016}  & \textbf{model} & \textbf{non-linear}\\
		\midrule
		CR & 1,559,923 & 0.551 & 0.659 & 0.658 & \bfseries 0.688 & 0.687\\
		CA & 1,212,948 & 0.537 & 0.673 & 0.673 & \bfseries 0.701 & 0.700\\
		ME & 6,443,490 & 0.546 & 0.664 & 0.664 & \bfseries 0.701 & 0.697\\
		NE &   802,070 & 0.539 & 0.643 & 0.643 & \bfseries 0.679 & 0.673\\
		SG & 3,249,435 & 0.573 & 0.749 & 0.749 & \bfseries 0.771 & 0.768\\
		\bottomrule
		\multicolumn{7}{l}{\scriptsize $^\ddagger$ \textbf{Cohort} cf. \cite{YaleNewHaven2019}: Cardiorespiratory (CR), Cardiovascular (CA), Medicine (ME),}\\
	\multicolumn{7}{l}{Neurology (NE) and Surgery/Gynaecology (SG)}
	\end{tabular}
\end{table}

\section{Conclusions}\label{sec:conclusion}

\textbf{Need for interpretability: }
In this work, we developed a neural network architecture for predicting health outcomes with {partial interpretability}: it allows us to extract the hospital-specific contribution to the overall health outcomes and, thereby, estimates the hospital-specific performance. It performs a so-called risk-adjustment, that is, it corrects for the different base populations of hospitals. Our work fills a gap in the existing body of research, as previous work relied upon linear models (\ie, hierarchical generalized linear models) and where thus any form of non-linearities and interaction effects among risk variables is discarded. As a remedy, our network architecture learns these effects and is capable of controlling for non-linearities. 

Even though our approach is subject to partial interpretability, this does not affect the overall prediction performance. To quantify this, the above analysis also compares our approach with partial interpretability against a non-interpretable network that operates fully as black-box (but can thus learn interactions between hospital dummies and risk variables). Despite the larger flexibility, the ROC-AUC from the latter approach is on par to our proposed architecture. Needless to say, only the partially-interpretable architecture allows for achieving our objective: estimating the risk-adjusted hospital performance under non-linearities. 

\textbf{New task/model: }
Our estimates differ from the tasks that are prevalent in healthcare \cite{Cheng2016, Che2016, Che2017, Ma2018}. Conventionally, machine learning for healthcare is interested in improving the prediction accuracy of health outcomes. For this, approaches from the realm of ``black-box'' modeling have become widespread. Different from that, it is imperative in our work that the approach is partially interpretable, as we must model hospital fixed-effects $\alpha_k$. Because of that reason, many machine learning models for healthcare cannot act as baselines and, instead, we compare our approach primarily against the hierarchical generalized linear model from \cite{YaleNewHaven2019} that is used in industry and represents the state-of-the-art. 

There are extensive theoretical arguments from medical research on why our non-linear approach is superior to the HGLM from industry practice. The rationale is two-fold. (1)~The influence of many risk factors may not be linear but, rather, follows a U-shaped relationship. Different from the HGLM, these non-linear relationships are directly considered by our proposed neural network. 
(2)~The influence of primary/secondary diseases is not additive. There might be interactions as some conditions are likely to co-occur, either because they are symptoms of the same cause or because their joint occurrence is worsening the overall health state. As an example, almost a quarter of patients with diabetes also suffer from a depression \cite{McDaniel1995}, where the progression of cancer affects that of depression and likely vice versa. These comorbidities are captured in our neural network architecture and their joint effect is automatically learned from the data. Both of the above reasons give a theoretical explanation of why the HGLM is outperformed by our proposed approach.  

\textbf{Implications for practice: }
Our work is of immediate relevance to decision-making in management and practice. For patients, our work provides findings that help them in choosing among hospitals. For policy-makers, we provide a tool for better monitoring of hospital performance. This is of direct value to government bodies that regularly compute risk-adjusted hospital performance. In the US, this includes the Hospital Readmissions Reduction Program, the {Centers for Medicare \& Medicaid}, and the {National Committee for Quality Assurance}.

We hope that our findings spur a discussion in healthcare management whether the current approach for performance monitoring is serving its purpose. Here we connect to a growing stream of literature that questions the current mechanism \cite{Finkelstein2016, Finkelstein2017} and make suggestions that further variables should be considered. In line with this, we provide evidence that the current approach in the form of the HGLM misses a large portion of variance in the outcomes that can be attributed to non-linear relationships. In that sense, our proposed neural network allows to quantify an upper bound of how much variance is missed due to a linear model. Hence, a direct policy implication is that the current approach should be expanded so that it is more effective in considering non-linearities among risk factors. Our proposed neural network provides a step in that direction. 

\section*{Acknowledgment}
The contribution of Patrick Z\"ochbauer in conducting this research is highly appreciated. 

\bibliographystyle{IEEEtran}
\bibliography{literature_abrv,literature}

\appendices

\section{Data}\label{app:data}
\noindent
\subsection{Pre-processing}
Our pre-processing is analogous to current practice for risk-adjustment of the 30-day readmission rate \cite{YaleNewHaven2019}. For motivations behind inclusion and exclusion criteria, we refer to \cite{Horwitz2011,YaleNewHaven2019}. In short, the following pre-processing steps are applied to our data:
\begin{enumerate}[leftmargin=0.5cm]
\item Patients who died or have non-existing primary diagnoses in the initial admission for an episode of care (\ie, the so-called \emph{index admission}) are removed from the dataset, as they can not be readmitted.
\item All ICD-10 diagnosis and procedure codes are mapped to Clinical Classification Software (CCS) diagnosis and procedure category codes, using the mapping of the Agency for Healthcare Research and Quality (AHRQ) \footnote{\url{www.hcup-us.ahrq.gov/toolssoftware/ccsr/ccs_refined.jsp}} \footnote{\url{www.hcup-us.ahrq.gov/toolssoftware/ccs10/ccs10.jsp}}.
\item Admissions with CCS diagnosis codes corresponding to psychiatry, cancer and rehabilitation are removed from the dataset \cite[cf. Table 3 of supplementary file]{YaleNewHaven2019}.
\item Readmissions are identified as admissions within 30 days after discharge of the index admission. Admissions are not considered readmissions when one the following conditions is met: if they are in the list of CCS diagnosis codes corresponding to psychiatry, cancer, and rehabilitation, if the procedure is planned \cite[cf. Table PR.1]{YaleNewHaven2019}; if the diagnosis was planned (\eg, rehabilitation or maintenance chemotherapy) \cite[cf. Table PR.2]{YaleNewHaven2019}; or if the procedure is potentially planned \cite[cf. Table PR.3]{YaleNewHaven2019} and if the diagnosis is not acute \cite[cf. Table PR.4]{YaleNewHaven2019}.
\item Hospitals with fewer than 500 admissions or no readmissions at all are removed from our dataset.
\item All ICD-10 diagnosis codes are mapped to ICD-9 codes. This was done in correspondence with the ICD version used in the embeddings of \cite{Choi2016} in line with procedures from by the Center for Medicare \& Medicaid Services (CMS) General Equivalence Mappings (GEM).\footnote{\url{www.cms.gov/Medicare/Coding/ICD10/2018-ICD-10-CM-and-GEMs}}
\item Patient age and gender are standardized.
\end{enumerate}

The following pre-processing steps have not been performed in comparison to \cite{YaleNewHaven2019}:
\begin{itemize}
\item Patient transfers from one health-care facility to another are not identified and therefore not excluded from our dataset.
\item Discharge against medical advice could not be identified and is therefore not excluded from our dataset.
\item Selections based on Medicare patient populations and selections based on characteristics of health-care facilities, such as facilities directed at acute care or facilities for cancer patients, could not be performed with our dataset.
\end{itemize}

\subsection{Train/test split}
The dataset was split into a training set (75\,\%), a validation set (18.65\,\%), and a test set (6.25\,\%). The splitting process is performed in stratified fashion with regard to hospital IDs, such that each set contains an equal amount of hospitals from the same ID. The prediction performance is measured out-of-sample based on the test set.\footnote{Note that our dataset covers only a single year and, hence, between-year trends cannot be present, because of which random splitting is used. We further refrained from using cross-validation due to the large size of the dataset (\ie, even the test set counts almost one million different observations). Nevertheless, we manually inspected the variance of the prediction performance across different splits, finding that it is extremely stable. This contributes to the robustness of our approach and further to the validity of our findings.}

\subsection{Descriptive statistics}
The dataset comprises 1,889 different hospitals. The majority of them (\ie, 64.8\,\%) is located in small metropolitan areas with less than 1 million inhabitants. 

Descriptive statistics for the data are reported in Table~\ref{tab:data-patient}. It summarizes admission data for the complete dataset. Overall, the dataset after preprocessing counts data from 13,267,866 admissions. The gender ratio is fairly balanced, while the age is around 60~years. This is representative of US hospitals. Overall, the 30-day readmission rate amounts to (\ie, 13.0\,\%) with a standard deviation of 4 percentage points. This indicates that there are significant differences between hospitals. 
In addition, Table~\ref{tab:data-patient} compares different subsets of the data, \ie, patients from five common disease cohorts as defined in \cite{YaleNewHaven2019}. 

\begin{figure}[t!]
	\centering
	\includegraphics[scale=.5, trim={.2cm .2cm .2cm .2cm}, clip]{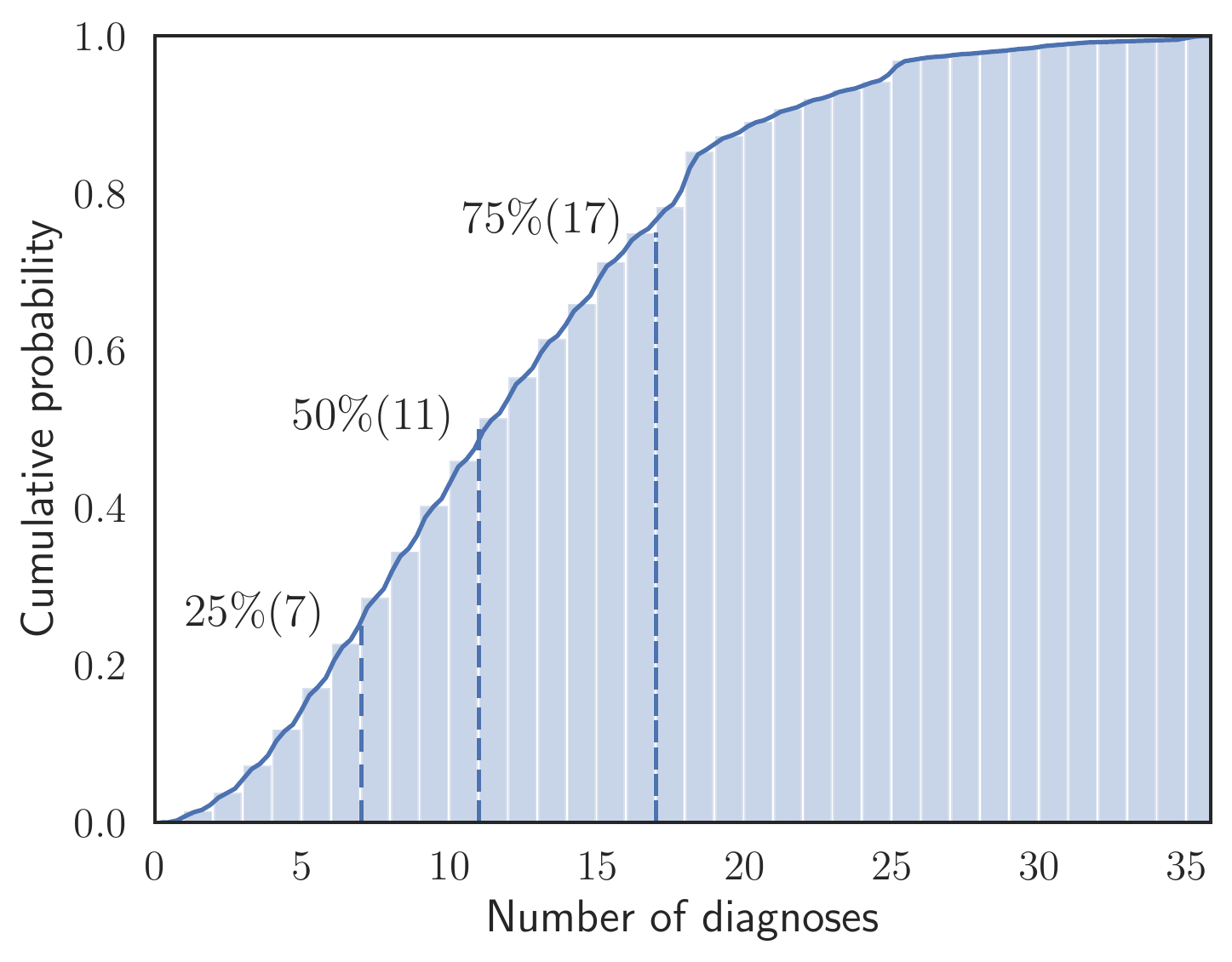}
	\caption{Cumulative distribution of diagnoses.}
	\label{fig:comorbidities}
\end{figure}

Table~\ref{tbl:top_icd} lists the most frequently occurring diagnoses. The most common one is high blood pressure, which is diagnosed in more than one-third of all hospitalizations. \figurename~\ref{fig:comorbidities} shows how the number of different diagnoses (\ie, $\abs{\bm{\phi}^i})$ is distributed among patients. Evidently, half of the patients have been assigned to 11 or more different diagnoses codes, with a maximum of 35. 

\begin{table}[t!]
	\centering
	\scriptsize
	\caption{Patient Descriptives Across Patient Cohorts and for the Complete Dataset}
	\label{tab:data-patient}
	\setlength{\tabcolsep}{3pt}
	\begin{tabular}{l r c c c}
		\toprule
		\textbf{Cohort}$^\parallel$ & \textbf{Admissions} & \textbf{Gender [\% f]} & \textbf{Age ($\sigma$)} &  \textbf{Readmission rate [\%]}\\
		\midrule
		CR  &   1,559,923 &   53.5 &      64.0 (21.2) &         17.5 \\
		CA  &   1,212,948 &   44.2 &      67.1 (15.0) &         12.9 \\
		ME  &   6,443,490 &   51.8 &      60.0 (20.7) &         15.4 \\
		NE  &     802,070 &   50.8 &      63.0 (20.4) &         11.8 \\
		SG  &   3,249,435 &   64.2 &      52.9 (20.4) &         6.6 \\
		\midrule
		\textbf{Total}              &  13,267,866 &   54.3 &      59.6 (20.7) &         13.0 \\
		\bottomrule
		\multicolumn{5}{l}{\scriptsize $^\parallel$ \textbf{Cohort} cf. \cite{YaleNewHaven2019}: Cardiorespiratory (CR), Cardiovascular (CA), Medicine (ME),}\\
		\multicolumn{5}{l}{Neurology (NE) and Surgery/Gynaecology (SG)}
	\end{tabular}
\end{table}

\begin{table}[t!]
	\centering
	\scriptsize
	\caption{Most Frequently Diagnosed Conditions}
	\label{tbl:top_icd}
	\setlength{\tabcolsep}{3pt}
	\begin{tabular}{lp{5.5cm} l}
		\toprule
		\textbf{ICD-10}  & \textbf{Description} & \textbf{Frequency} \\  \midrule
		I10  	& High blood pressure 									& 5,559,365\\
		E78.5 	& Hyperlipidemia, unspecified 							& 4,015,345\\
		K21.9  	& Gastroesophageal reflux disease without esophagitis 	& 2,600,654\\
		Z87.891 & Personal history of nicotine dependence 				& 2,494,972\\
		I25.10  & Atherosclerotic heart disease of native coronary artery without angina pectoris &2,487,193\\
		E11.9  	& Type 2 diabetes mellitus without complications 		& 1,928,030\\
		N17.9 	&  Acute kidney failure, unspecified 					& 1,797,255\\
		E03.9  	&  Hypothyroidism, unspecified 							& 1,643,776\\
		F17.210 & Nicotine dependence, cigarettes, uncomplicated 		& 1,595,312\\
		Z79.82 	& Long term (current) use of aspirin 					& 1,556,844\\
		\bottomrule
	\end{tabular}
\end{table}

\section{Model}\label{app:model}
\subsection{Learning algorithm}
Our neural network is trained using the algorithm described in Section~\ref{sec:estimation-details}, with a batch size of 16,384.\footnote{The batch size is optimized for a fixed architecture with parameter options: $\{1024, 2048, 4096, 8192, 16384, 32768\}$.} The network is optimized until the loss on the validation data has stopped improving, with a patience of five training epochs and it returns the best performing model evaluated on the validation data. The algorithm uses cyclical learning rates \cite{Smith2017} with a triangular learning rate policy. For the triangular learning rates, a step size of $10^3$ is used, corresponding to two times the number of training iterations in one epoch. The initial learning rate is set to $1\times 10^{-3}$ and the upper boundary of the learning rate is set to $6\times 10^{-3}$.

To account for imbalances in the dataset, the model is estimated using a class-weighted loss function, such that weights are updated inversely proportional to the class imbalance. Additionally, the loss function is subject to $L_2$-regularization as described in (\ref{eq:loss}). Here we applied a shrinkage parameter of $\lambda = 10^{-5}$. The $L_2$-regularization shrinkage parameter was optimized for a fixed architecture across the following range: $\{10^{-3}, 10^{-4}, 10^{-5}, 10^{-6}\}$.

For practical applications, our proposed approach achieves a favorable computational runtime. In fact, the runtime remains below 2~hours on standard server architecture. This must be viewed in the context that in the US, only one such analysis is currently performed per year.

\subsection{Hyperparameter tuning}
The model hyperparameters are optimized using a grid search. The following hyperparameters of the proposed model are optimized: (i)~the number of dense hidden layers that follow patient diagnoses information ($L_q$); (ii)~the number of dense hidden layers that follow all patient-specific information ($L_p$); (iii-iv)~for each group of dense hidden layers (\ie, $L_q$ or $L_p$) the number of nodes per layer and (v)~the percentage of nodes that is dropped after each hidden layer. The hyperparameter-space for our proposed model is listed in Table~\ref{tab:hyp}. The hyper-parameters that obtain the best performance measured with ROC-AUC are underlined.
 
Additionally, some hyperparameters are fixed and they are initialized as follows. The activation function of each dense hidden layer is taken to be ReLU. The bias initialization in the last layer of the network is initialized at the average health outcome of the training set.

For a fair comparison of model performance, the hyperparameters of the HGLM and Elastic net baselines (\ie, $L_1$-regularization shrinkage $\lambda_1$ and $L_2$-regularization shrinkage $\lambda_2$) are optimized in similar fashion as described above. The hyperparameter grid for the baselines is the following, whereby the model with the highest ROC-AUC evaluated on the validation set is underlined:
\begin{itemize}
	\item \textit{HGLM}: $\lambda_2 = \{10^{-3}, 10^{-4}, 10^{-5}, \underline{10^{-6}}\}$;
	\item \textit{Elastic net}: $\lambda_1 = \{10^{-3}, 10^{-4}, 10^{-5}, \underline{10^{-6}}\}$ and\\ $\lambda_2 = \{10^{-3}, 10^{-4}, 10^{-5}, \underline{10^{-6}}\}$.
\end{itemize}

\begin{table}[t!]
	\centering
	\scriptsize
	\caption{Tuning Grid for Hyperparameters}
	\label{tab:hyp}
	\begin{tabular}{l l}
		\toprule
		\textbf{Layer} & \textbf{Options}\\
		\midrule
		Number of layers $L_q$ & $\{\underline{0},1,2\}$\\
		Number of layers $L_p$ & $\{0,1,\underline{2}\}$\\
		Number of nodes per layer $L_q$ & $\{\underline{512}, 1024\}$\\
		Number of nodes per layer $L_p$ & $\{\underline{256}, 512\}$\\
		Dropout after each layer & $\{\underline{25\,\%}, 50\,\%\}$ \\
		\bottomrule
		\multicolumn{2}{l}{
			\begin{minipage}{.32\textwidth}
				\begin{itemize}[leftmargin=*, align=parleft, labelsep=1.1cm]
					\item[\underline{underlined}] Values corresponding to the best performance
				\end{itemize}	
			\end{minipage}
		}
	\end{tabular}
\end{table}

\subsection{Prediction performance}
\begin{figure}[t!]
	\centering
	\includegraphics[scale=.5, trim={0.2cm 0.2cm 0.2cm 0.8cm}, clip]{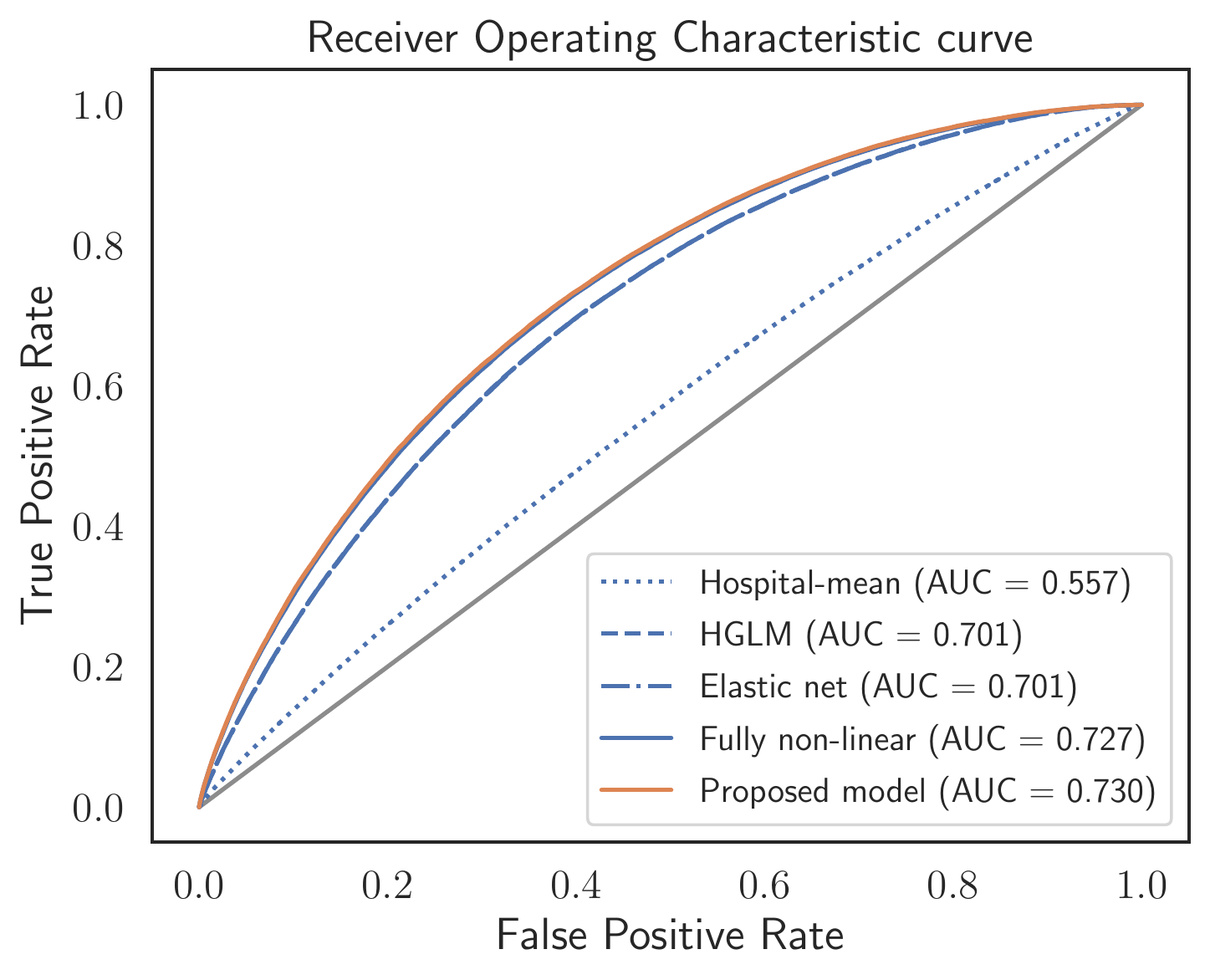}
	\caption{Receiver operating characteristic (ROC) curve.}
	\label{fig:roc}
\end{figure}

\begin{figure}[t!]
	\centering
	\includegraphics[scale=.5, trim={0.2cm 0.2cm 0.2cm 0.8cm}, clip]{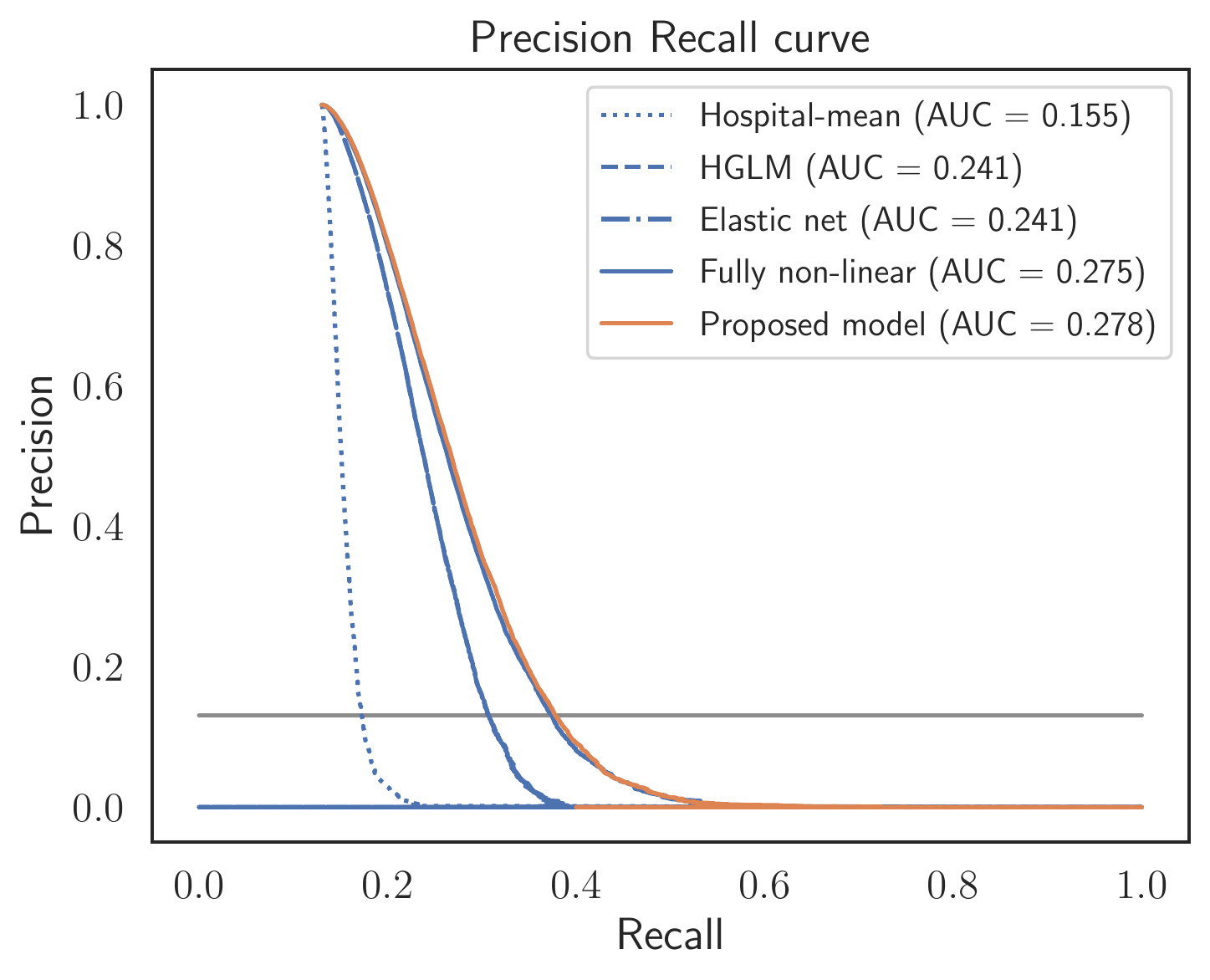}
	\caption{Precision-recall curve.}
	\label{fig:pr}
\end{figure}

The prediction performance is further detailed as follows. (i)~\figurename~\ref{fig:roc} shows the ROC curve, thus comparing the false positive rate against the true positive rate. (ii)~\figurename~\ref{fig:pr} presents the precision-recall curve which is common for unbalanced data. Both figures confirm that our proposed model has superior performance. 

\end{document}